\documentclass [12pt]{article}

\usepackage{amssymb,amsmath, amsthm}
\usepackage{graphicx, color,mathrsfs}
\usepackage{latexsym}
\usepackage{amsfonts}
\usepackage{amscd}
\usepackage{txfonts}
\usepackage[all]{xy}
\usepackage{float}

 \newtheorem{thm}{Theorem}[]
 
 \newtheorem{lem}[thm]{Lemma}
 \newtheorem{prop}[thm]{Proposition}
 \theoremstyle{definition}
 \newtheorem{defn}[thm]{Definition}
 \theoremstyle{remark}
 
\newcommand{\be}{\begin{equation}}
\newcommand{\ee}{\end{equation}}
\newcommand{\beq}{\begin{eqnarray}}
\newcommand{\eeq}{\end{eqnarray}}
\newcommand{\dis}{\displaystyle}

\newcommand{\la}{\label}

\newfont{\msbm}{msbm10 scaled\magstep1}
\newfont{\msbms}{msbm7 scaled\magstep1} 
\newif\ifintrmk
\newcommand{\intremark}[1]{\ifintrmk\par\bigskip\noindent\hrulefill\par\medskip \noindent{\bf Internal Remark:}
#1 \hfill$\clubsuit$\par\medskip\noindent\hrulefill\par\bigskip\else\fi}
\intrmkfalse           

\begin{document}
\begin{center}
{\LARGE{Absolutely Continuous Spectrum for the Anderson Model on Some Tree-like Graphs}}
\end{center}

\begin{center}
{\Large {Florina Halasan}} \footnote{Email address: florina.halasan@gmail.com\\  The author was supported by an NSERC/MITACS Scholarship. This paper is based on part of the author's doctoral thesis.\\ It is a pleasure to thank Dr. R. Froese for all his guidance and support.}
\end{center}

\begin{center}
\emph{Department of Mathematics, University of British Columbia, \\1984 Mathematics Road, Vancouver, B.C. V6T 1Z2 Canada}
\end{center}

\begin{abstract} We prove persistence of absolutely continuous spectrum for the Anderson model on a general class of tree-like graphs.
\end{abstract}



\section{Introduction}

Random Schr{\"o}dinger Operators are used as models for disordered
quantum mechanical systems. In particular, the Anderson Model was introduced to describe the
motion of a quantum-mechanical electron in a crystal with
impurities. For this model, the states corresponding to an absolutely
continuous spectrum describe mobile electrons. Thus, an interval of
absolutely continuous spectrum is an energy range in which the
material is a conductor.

An outstanding open problem, the extended states conjecture, is to
prove existence of absolutely continuous spectrum for the lattice
$\mathbb{Z}^d$ with $d>2$. Until now, it is only for the Bethe lattice
that this has been established. A first result on the topic was
obtained by A. Klein,~\cite{K}, in $1998$; he proved that for weak
disorder, on the Bethe lattice, there exists absolutely continuous
spectrum for almost all potentials. More recently, Aizenman, Sims
and Warzel proved similar results for the Bethe lattice
using a different method (see~\cite{ASW}). Their method establishes the persistence
of absolutely continuous spectrum under weak disorder and also in the presence of a
periodic background potential. During the same time, Froese, Hasler and Spitzer introduced a geometric method for proving the existence of absolutely continuous spectrum on graphs (see~\cite{FHS-1}). In their second paper on the topic,~\cite{FHS-2}, they proved delocalization for the Bethe lattice of degree $3$ using this geometric approach. 

In this work, we provide a version of the geometric method on a more general class of trees.

\subsection*{Statement of the Main Result}

\begin{figure}[H]
\begin{center}
\includegraphics[scale=0.6]{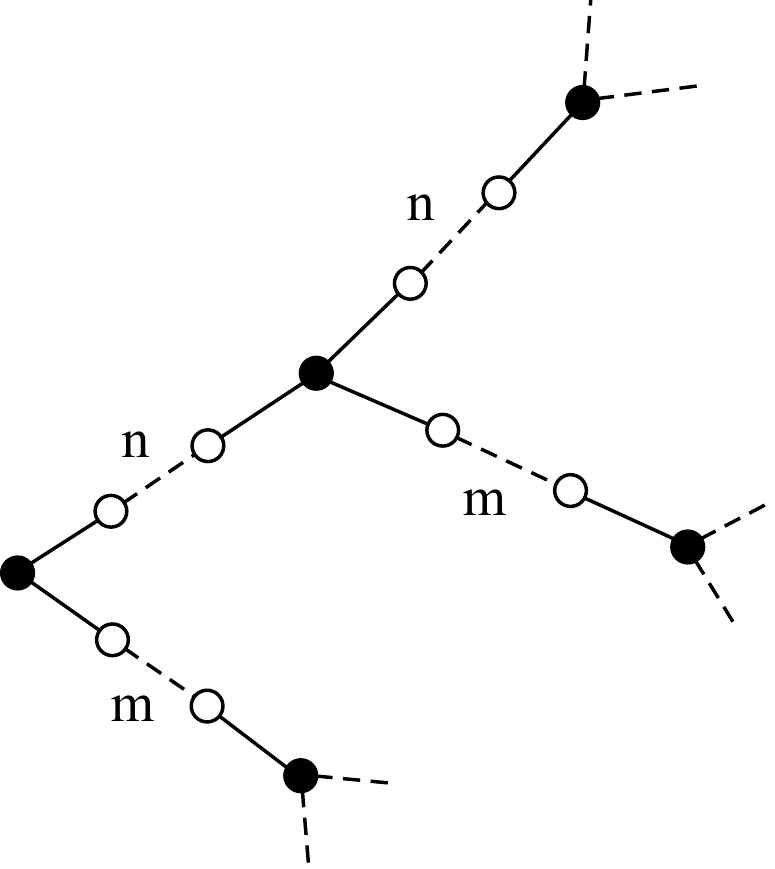}
\caption{The $\mathbb{T}$ tree} \label{Fig.1}
\end{center}
\end{figure}

We prove the existence of purely absolutely continuous spectrum for the
Anderson Model on a tree-like graph, $\mathbb{T}$, defined as follows 
(see Figure \ref{Fig.1}). 

\begin{defn} Let $\mathbb{B}$ be an infinite full binary tree in which each node has degree $3$ except for the origin, which has degree $2$.  Let us call its nodes principal nodes and denote by $o$ its origin. For the origin and each principal node there are two edges leading away from the origin. Choose one of them and call it the top edge and call the other the bottom edge. On each top edge, we add $m$ distinct auxiliary nodes; similarly we add $n$, $m \neq n$, distinct auxiliary nodes on each bottom edge. Thus we obtain the tree $\mathbb{T}$ which has a set of principal nodes denoted by $\mathbb{T}_p$ and a set of auxiliary nodes denoted by $\mathbb{T}_a$.
\end{defn}

The conclusions in this paper remain valid if we start with any k-nary tree. We present the binary case for simplicity. By excluding the $m=n$ case we break some of the symmetry in our tree; this asymmetry is used in Proposition 4, Section 3. The proof for the $m = n$ case would constitute a generalization of the Bethe lattice proof presented in~\cite{H} and be considerably longer.  

Using the terminology established in~\cite{ASW}, we will use the symbol $\mathbb{T}$ for both our tree graph and its set of vertices. For each $x \in \mathbb{T} =\{o\} \cup \mathbb{T}_p \cup \mathbb{T}_a$ we have at most one neighbor towards the root and two in what we refer to as the forward direction. We say
that $y \in \mathbb{T}$ is in the future of $x \in \mathbb{T}$ if the path
connecting $y$ and the root runs through $x$. The subtree consisting
of all the vertices in the future of $x$, with $x$ regarded as its
root, is denoted by $\mathbb{T}^x$.

The Anderson Model on $\mathbb{T}$ is given by the random Hamiltonian, $H$,
on the Hilbert space $\ell^2(\mathbb{T})\;=\;\Big\{\;\varphi: \mathbb{T} \rightarrow \mathbb{C} \,;\;\sum\limits_{x\in \mathbb{T}} {\left| {\varphi(x)} \right|^2<\infty }\Big\}$. This operator is of the form
\begin{equation*} H = \Delta + k \,q \,\nonumber \end{equation*}
where:
\begin{enumerate}

 \item The free Laplacian $\Delta$ is defined by
 $$(\Delta \varphi)(x)=\sum_{y:d(x,y)=1}\left(\varphi(x)-\varphi(y)\right)
 \,, \,\rm{for \, all}\, \varphi \in \ell^2(\mathbb{T})\,,$$
where the distance $d$ denotes the number of edges between sites.

\item The operator $q$ is a random potential,
$$(q\varphi)(x)=q(x)\varphi(x),$$
where $\{q(x)\}_{x\in\mathbb{T}}$ is a family of independent, identically distributed real random variables
with common probability distribution $\nu$. We assume the $2(1+p)$ moment,\\ $\int |q|^{2(1+p)}d\nu$, is finite for some $p>0$. The coupling constant $k$ measures the disorder.
\end{enumerate}
Our main theorem states that the above defined Anderson model exhibits purely absolutely continuous spectrum for low disorder. 
\smallskip

\begin{thm} \la{Main} Let $F$ be the open interior of the absolutely continuous spectrum of $\Delta$ (this spectrum depends on $m$ and $n$) with a finite set of values, $S$, removed. For any closed subinterval $E$, $ E \subset F$, there exists $k(E) > 0$ such that for all $0<|k|<k(E)$ the spectrum of $H$ is purely absolutely continuous in $E$ with probability one.
\end{thm}

\noindent {\it Remarks.}
\newline $1).$ The finite set $S$ will be properly identified in  Proposition \ref{P3.1}.
\newline $2).$ The actual definition we use for $F$ is $F :=\{ \lambda \in \mathbb{R}:  z_{\lambda} \in \mathbb{C},  \mathrm{Im}(z_\lambda)>0\}\setminus S$ where $z_\lambda= \langle \delta_o, (\Delta-\lambda)^{-1}\delta_o \rangle$ ($\delta_o$ is the indicator function at the origin). Defined like this, $F$ is the support of the absolutely continuous component of the spectral measure of $\Delta$ for $\delta_o$ without the special values contained in $S$. Following the ideas in Lemma \ref{Lx} (Section~\ref{appA}), i.e. rearranging the tree and deriving a formula for the Green function at the new origin, we can prove that the set $\{ \lambda \in \mathbb{R}:  z_{\lambda} \in \mathbb{C},  \mathrm{Im}(z_\lambda)>0\}$ is, in fact, the support of the pure absolutely continuous spectrum for the Laplacian $\Delta$.


\noindent  Let $\delta_x  \in \ell^2(\mathbb{T})$ be the indicator function supported at the site $x \in \mathbb{T}$ and let $R(E,\epsilon) = \{z\in \mathbb{C} : {\rm Re}(z) \in E, 0<{\rm Im}(z) \leq \epsilon \}$ be a strip along the real axis, for $E$ defined in the previous theorem. The following theorem together with the criterion from Section~\ref{appA} gives us the proof of Theorem \ref {Main}.

\begin{thm} \la{T3.2}Under the hypothesis of the previous theorem, we have
$$
\sup_{\lambda \in R(E,\epsilon)} \mathbb{E} \left( \left|\left\langle {\delta_x,(H -\lambda)^{-1}\delta_x} \right\rangle\right|^{1+p} \right) < \infty \; ,
$$
for all sufficiently small $p>0$, some $\epsilon > 0$ and all $x \in \mathbb{T}$. 
\end{thm}

\noindent{\bf Proof of Theoreom \ref{Main}.} Let us consider $\lambda = \alpha +{\rm i} \beta$. Using Fatou's lemma, Fubini's theorem and Theorem \ref{T3.2}, we obtain
\begin{align*}
&\mathbb{E} \left( \liminf_{\beta \searrow 0} \int_E  \left| \langle \delta_x ,
(H-\lambda )^{-1}   \delta_x \rangle \right|^{1+p} d\alpha  \right)\\
&\leq  \liminf_{\beta \searrow 0} \int_E \mathbb{E} \left( \left| \langle \delta_x, (H-\lambda )^{-1}  \delta_x \rangle \right|^{1+p}  \right) d\alpha< \infty \,. 
\end{align*}
Therefore we must have
\begin{eqnarray*}
\liminf_{\beta \searrow 0} \int_E  \left| \langle \delta_x ,(H-\lambda
)^{-1} \delta_x \rangle \right|^{1+p} d\alpha < \infty\,,
\end{eqnarray*}
with probability one. Since  $\langle \delta_x ,(H-\lambda )^{-1} \delta_x\rangle$ is the Stieltjes transform of the measure
$d\mu_x$, it follows from Proposition \ref{Simon}, Section 5 that the restriction of $\mu_x$ 
to $E$ is purely absolutely continuous with probability one. In other words, the spectral measure for $H$ corresponding to $\delta_x$, for any $x \in \mathbb{T}$, is purely absolutely continuous in $E$ with probability one. Therefore the operator $H$ has purely absolutely continuous spectrum on $E$.
\qed

\section{Outline of the Proof}

Let $G_x(\lambda)=\left\langle {\delta_x,(H -\lambda)^{-1}\delta_x}\right\rangle$ denote the diagonal matrix element of the resolvent at some arbitrary vertex $x \in \mathbb{T}$, often referred to as the Green function. Our goal is to find bounds for these Green functions. We first do so for $G_o(\lambda)$ and then extend the bound to all diagonal terms.

Let $H^x$ be the restriction of $H$ to $\ell^2(\mathbb{T}^x)$. The forward Green function $G^x(\lambda)$ is defined to be the Green function for the truncated graph, given by
\begin{equation*}
G^x(\lambda)=\left\langle {\delta_x,(H^x -\lambda)^{-1}\delta_x}
\right\rangle \;.
\end{equation*} 
\begin{figure}[H]
\begin{center}
\includegraphics[scale=0.7]{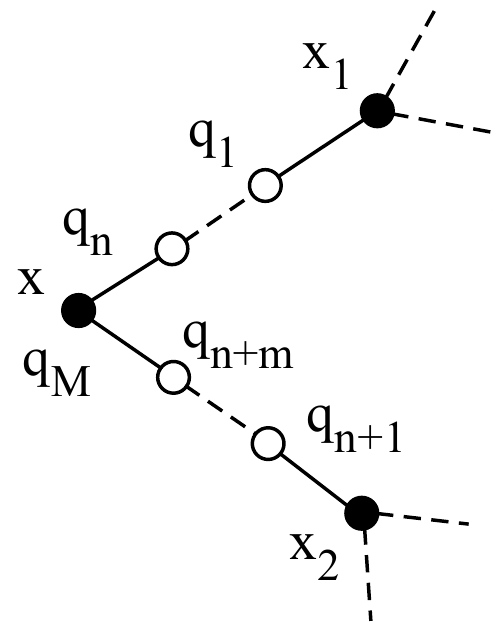}
\caption{The nodes in the recurrence relation for the forward Green function.} \label{Fig.2}
\end{center}
\end{figure}
\noindent The forward Green function $G^x(\lambda)$, for $x$ a principal node, can be expressed recurrently as a function depending on the forward Green function for the two forward principal nodes and all the random potentials in between. Thus the recurrence relation, which can be derived using resolvent properties, has the form
\begin{equation}\label{gf_rec}
 G^x(\lambda)=\phi(G^{x_1}(\lambda),G^{x_2}(\lambda),q_1 \ldots q_M,\lambda),
 \end{equation}
where
$$
\phi:\mathbb{H}^{2}\times \mathbb{R}^M \times \mathbb{H} \to \mathbb{H}
$$
is defined by
\begin{equation} 
\phi(z_1,z_2,q_1 \ldots q_M,\lambda) =
    \frac{-1}{\phi_n(z_1,q_1\ldots q_n,\lambda)+\phi_m(z_2,q_{n+1}
    \ldots q_{n+m},\lambda)+\lambda-q_M}\,  \la{rec}
\end{equation}
with
\begin{eqnarray*}
  \phi_0(z,\lambda)&=& z  \\
  \phi_1(z,q_1,\lambda)& = &\frac{-1}{z+\lambda-q_1+1}   \\
  \ldots \\
  \phi_n(z,q_1 \ldots q_n,\lambda)& = &
    \frac{-1}{\phi_{n-1}(z,q_1 \ldots q_{n-1},\lambda)+\lambda-q_n+1}   \,
\end{eqnarray*}
and $M=m+n+1$.
The nodes $x,x_1,x_2$  and the potentials $q_1,\cdots,q_M$ involved in the recurrence \eqref{gf_rec} are shown in  Figure \ref{Fig.2}. Because the origin has degree 2, the recurrence relation for $G^o(\lambda)$ is given by $G^o(\lambda)=\phi(G^{x_1}(\lambda),G^{x_2}(\lambda),q_1 \ldots q_M,\lambda+1)$.

In the above definition $\mathbb{H} = \{z \in \mathbb{C}: \mathrm{Im}(z) >0 \}$ is the complex upper half plane. Notice that since $H$ is random, at each $x \in \mathbb{T}$, the forward Green
function $G^x(\lambda)$ is an $\mathbb{H}$-valued random variable. We notice that since the random potential is i.i.d., at any $x \in \mathbb{T}_p$, $G^x(\lambda)$ has the same probability distribution denoted by  $\rho$. The Green function at the origin, $G_o(\lambda) = G^o(\lambda)$, has probability distribution denoted by $\rho_o$.

The transformations $\phi_n$ and $\phi_m$, in the recursion formula, are compositions of fractional linear transformations, hence fractional linear transformations themselves. This implies $\phi$ is a rational function whose numerator and denominator have degree $2$. If ${\rm Im}( \lambda) > 0$, the map $z \mapsto \phi(z,z,0,\ldots,0,\lambda)$ is an analytic map from $\mathbb{H}$ to $\mathbb{H}$, a hyperbolic contraction. Let $z_\lambda$ denote its unique fixed point in the upper half plane, a solution to the cubic equation $z =\phi(z,z,0,\ldots,0,\lambda)$. The set  $\{ \lambda: {\rm Im}(\lambda)=0\,\, {\rm and}\,\, z_\lambda \in \mathbb{H}\}$ is a reunion of two disjoint open intervals on the real axis. Thus, for $q \equiv 0$, $G^x(\lambda)=z_\lambda$ for all $x \in \mathbb{T}_p$ and $G^o(\lambda)=\phi(z_\lambda,z_\lambda,0 \ldots 0,\lambda+1)$. The map $\mathbb{H}\ni\lambda \mapsto z_\lambda$ extends continuously onto the real axis. Therefore we define $F:=\{ \lambda: {\rm Im}(\lambda)=0\,\, {\rm and}\,\, z_\lambda \in \mathbb{H}\}\setminus S$ where, as mentioned before, $S$ is defined in Proposition \ref{P3.1}. We should note again that the set $F \cup S$ is the support of the absolutely continuous component of the spectral measure for $\delta_o$, for the free Laplacian. The set $\{z_\lambda\}_{\lambda\in E}$ is a compact curve strictly contained in $\mathbb{H}$. Thus, when $\lambda$
lies in the strip
$$
R(E,\epsilon) = \{z\in \mathbb{H} : {\rm Re}(z) \in E, 0<{\rm Im}(z) \leq
\epsilon \}
$$
with $E \subset F$ closed and $\epsilon$ sufficiently small, ${\rm Im}(z_\lambda)$ is bounded
below and $|z_\lambda|$ is bounded above.

To prove absolutely continuous spectrum, we need the bound on $|G_x|^{1+p}$ stated in Theorem 2. To get this bound we first prove that $\mathrm{w}^{1+p}(G_x)$ is bounded, where $\mathrm{w}$ is a weight function defined as follows:
\begin{equation*}
\mathrm{w}(z) = 2 (\cosh({\rm
dist}_{\mathbb{H}}(z,z_\lambda))-1) = \frac{|z-z_\lambda|^2}{{\rm
Im}(z){\rm Im}(z_\lambda)} \; .
\end{equation*}
Up to constants, $\mathrm{w}(z)$ is the hyperbolic cosine of the hyperbolic distance from $z$ to $z_\lambda$, the Green function at the root for $\Delta$. We have dropped the $\lambda$-dependence from the notation. 

Our proof relies on a pair of lemmas about the following quantity:
\begin{eqnarray*}
\mu_p(z_1,z_2,q_1\ldots q_M, \lambda) = \frac{\mathrm{w}^{1+p}(\phi(z_1,z_2,q_1\ldots q_M,\lambda))+
\mathrm{w}^{1+p}(\phi(z_2,z_1,q_1\ldots q_M,\lambda))}{ \mathrm{w}^{1+p}(z_1)
+ \mathrm{w}^{1+p}(z_2) } \,,
\end{eqnarray*}
for $z_1, z_2 \in \mathbb{H}^2$, $q_1,\ldots,q_M \in \mathbb{R}$ and $\lambda \in R(E,\epsilon)$.   
\begin{lem} \la{L3.1} For any closed subinterval $E$, $E \subset F$ and  all sufficiently small
$0<p<1$, there exist positive constants $\epsilon$, $\eta_1$,
$\epsilon_0$ and a compact set ${\cal K} \subset \mathbb{H}^2$ such that
\begin{eqnarray}
  \mu_p|_{{\cal K} ^c \times [-\eta_1, \eta_1]^M \times R(E,\epsilon_0 )}(z_1, z_2, q_1\ldots q_M, \lambda) &\leq& 1- \epsilon .
\end{eqnarray}
Here ${\cal K} ^c$ denotes the complement $\mathbb{H}^{2} \setminus {\cal
K}$.\end{lem}

\begin{lem} \la{L3.2} For any closed subinterval $E$, $E\subset F$ and any $0<p<1$, there exist positive constants $\epsilon_0$, $C$ and a compact set ${\cal K} \subset
\mathbb{H}^2$ such that
\begin{eqnarray}
  \mu_p|_{{\cal K}^c \times \mathbb{R}^M \times R(E,\epsilon_0 )}(z_1, z_2, q_1\ldots q_M, \lambda)
  &\leq& C \prod\limits_{i=1}^M (1+ |q_i|^{2(1+p)}) \, .
\end{eqnarray}
\end{lem}

Given these two lemmas we can prove that the decay of the probability distribution function
of the forward Green function at infinity is preserved as ${\rm
Im}(\lambda)$ becomes small, provided that $\nu$ has a finite moment of order $2(1+p)$. Using Lemma \ref{L3.1} and Lemma \ref{L3.2} we prove Theorem \ref{T3.3} below, the last ingredient needed in the proof. 
\vspace{0.3cm}

\begin{thm} \la{T3.3} For any closed subinterval $E$, $ E \subset F$, there exists $k(E) > 0$ such that for all $0<|k|<k(E)$ we have
$$
\sup_{\lambda \in R(E,\epsilon)} \mathbb{E} \left(  \mathrm{w}^{1+p}(G^x(\lambda)) \right)
 < \infty \; ,
 $$
 for all $x \in \mathbb{T}_p$.
 \end{thm}
 
\proof Let $\eta_1$ and $p$ be given by Lemma \ref{L3.1}, and
choose $\epsilon_0$ and $\cal K$ that work in both Lemma \ref{L3.1}
and Lemma \ref{L3.2}. For any $(z_1,z_2)\in {\cal K}^c$ and $\lambda
\in R(E,\epsilon)$, we estimate
\begin{align*}
&\int_{\mathbb{R}^M}\mu_p(z_1,z_2, k q_1\ldots k q_M,\lambda)
d\nu(q_1) \ldots d\nu(q_M) \\
&\leq
(1-\epsilon)\int_{[-\frac{\eta_1}{k},\frac{\eta_1}{k}]^M}d\nu(q_1)\ldots
d\nu(q_M)+C\int_{\mathbb{R}^M
\backslash[-\frac{\eta_1}{k},\frac{\eta_1}{k}]^M}\prod_{i=1}^M\left( 1+
|k\,q_i|^{2(1+p)}\right)d\nu(q_1)\ldots \\&\,\,\,\,\,\,\,\,\,\,\,\,\,\,\,\,\,\,\,\,\,\,\,\,\,\,\,\,\,\,\,\,\,\,\,\,\,\,\,\,\,\,\,\,\,\,\,\,\,\,\,\,\,\,\,\,\,\,\,\,\,\,\,\,\,\,\,\,\,\,\,\,\,\,\,\,\,\,\,\,\,\,\,\,\,\,\,\,\,\,\,\,\,\,\,\,\,\,\,\,\,\,\,\,\,\,\,\,\,\,\,\,\,\,\,\,\,\,\,\,\,\,\,\,\,\,\,\,\,\,\,\,\,\,\,\,\,\,\,\,\,\,\,\,\,\,\,\,\,\,\,\,\,\,\,\,\,\,\,\,\,\,\,\,\,\,\,\,\,\,\,\,\,\,d\nu(q_M) \\
&\leq 1-\epsilon/2, 
\end{align*}
provided $|k|$ is sufficiently small. 

The probability distributions on the hyperbolic plane are defined by
\begin{align*}
&\rho(A) = {\rm Prob} \{ G^x(\lambda) \in A \} \,,
\end{align*}
where $x$ is any site in $\mathbb{T}_p$. The recursion formula for the Green function implies that the distributions $d\rho$ 
are related by
\begin{align*}
\rho(A) &= {\rm Prob}\{\phi(z_1, z_2,k\,q_1\ldots k\,q_M,\lambda)\in
A \} = {\rm Prob}\{(z_1, z_2,k\,q_1 \ldots k\,q_M,\lambda)\in
\phi^{-1}(A) \}  \, \\ &=\int_{\phi^{-1}(A)}
\,d\rho(z_1)\,d\rho(z_2)\,d\nu(q_1)\ldots d\nu(q_M)) \,
\\ &=\int_{\mathbb{H}^2\times\mathbb{R}^M} \chi_A (\phi(z_1,
z_2,k\,q_1\ldots k\,q_M,\lambda))\,d\rho(z_1)\, d\rho(z_2)\,d\nu(q_1)\ldots
d\nu(q_M)
\end{align*}
which gives us that for any bounded continuous function $f(z)$
$$
\int_{\mathbb{H}}f(z) d\rho(z) = \int_{\mathbb{H}^2\times\mathbb{R}^M}
f(\phi(z_1,z_2,k\,q_1 \ldots k\,q_M,\lambda))\,d\rho(z_1)\,d\rho(z_2)\,d\nu(q_1)
\ldots  d\nu(q_M).
$$
Using this relation, for $\lambda \in R(E,\epsilon_0)$, we obtain
\begin{align*}
&\mathbb{E}\left(\mathrm{w}^{1+p}(G^x(\lambda)) \right) = \int_{\mathbb{H}} \mathrm{w}^{1+p}(z)\, d \rho(z) \\
&= \int_{\mathbb{H}^2\times\mathbb{R}^M} \mathrm{w}^{1+p}(\phi(z_1,z_2,k\,q_1 \ldots k\,q_M,\lambda))\,  d\rho(z_1)\,d\rho(z_2)\, d\nu(q_1) \ldots  d\nu(q_M) \\
&=\int_{\mathbb{H}^2\times\mathbb{R}^M} \frac{1}{2} \big( \mathrm{w}^{1+p}(\phi(z_1,z_2,k\,q_1 \ldots k\,q_M,\lambda))+ \mathrm{w}^{1+p}(\phi(z_2,z_1,k\,q_1 \ldots k\,q_M,\lambda))\big)\,
d\rho(z_1) \, \\ &\,\,\,\,\,\,\,\,\,\,\,\,\,\,\,\,\,\,\,\,\,\,\,\,\,\,\,\,\,\,\,\,\,\,\,\,\,\,\,\,\,\,\,\,\,\,\,\,\,\,\,\,\,\,\,\,\,\,\,\,\,\,\,\,\,\,\,\,\,\,\,\,\,\,\,\,\,\,\,\,\,\,\,\,\,\,\,\,\,\,\,\,\,\,\,\,\,\,\,\,\,\,\,\,\,\,\,\,\,\,\,\,\,\,\,\,\,\,\,\,\,\,\,\,\,\,\,\,\,\,\,
d\rho(z_2)\, d\nu(q_1) \ldots d\nu(q_M)\\
&= \frac{1}{2} \int_{{\cal K}^c}\left(\int_{\mathbb{R}^M}
\mu_p(z_1,z_2,k\,q_1 \ldots k\,q_M,\lambda)\,d\nu(q_1)\,\ldots\,
d\nu(q_{M})\right) \times \big( \mathrm{w}^{1+p}(z_1)  + \mathrm{w}^{1+p}(z_2) \big)\\&\,\,\,\,\,\,\,\,\,\,\,\,\,\,\,\,\,\,\,\,\,\,\,\,\,\,\,\,\,\,\,\,\,\,\,\,\,\,\,\,\,\,\,\,\,\,\,\,\,\,\,\,\,\,\,\,\,\,\,\,\,\,\,\,\,\,\,\,\,\,\,\,\,\,\,\,\,\,\,\,\,\,\,\,\,\,\,\,\,\,\,\,\,\,\,\,\,\,\,\,\,\,\,\,\,\,\,\,\,\,\,\,\,\,\,\,\,\,\,\,\,\,\,\,\,\,\,\,\,\,\,\,\,\,\,\,\,\,\,\,\,\,\,\,\,\,\,\,\,\,d\rho(z_1) \,  d\rho(z_2) + C \\
&\leq ( 1 - \epsilon/2) \int_{\mathbb{H}} \mathrm{w}^{1+p}(z)\, d \rho(z) + C = ( 1 - \epsilon/2)\, \mathbb{E}\left(\mathrm{w}^{1+p}(G^x(\lambda)) \right) +C,
\end{align*}
where $C$ is some finite constant, only depending on the choice of
$\cal K$. This implies that for all $\lambda \in R(E,\epsilon_0)$,
$$
\mathbb{E} \left( \mathrm{w}^{1+p}( G^x(\lambda) )  \right) \leq \frac{2C}
{\epsilon}  \; .
$$
\qed

\bigskip
\noindent {\bf Proof of Theorem 2.} It is an immediate consequence of Theorem \ref{T3.3}, Lemma  \ref{Lx} and the following inequality which holds for any two complex numbers $z$ and $s$ in $\mathbb{H}$:
\begin{equation} \la{ineq:abs3}
|z| \leq 4 {\rm Im}(s) \frac{|z-s|^2}{{\rm Im}(z){\rm Im}(s)} + 2 |s| \; .
\end{equation}
The inequality clearly holds for $|z| \leq 2 |s|$. In the
complementary case, we have $|z| > 2 |s|$ and thus $|z-s| \geq
||z|-|s|| \geq |s|$, implying
$$
|z|{\rm Im}(z) \leq |z|^2 \leq 2 \left( |z-s|^2 +  |s|^2 \right)\leq 4 |z - s|^2
\;
$$
and further $|z| \leq 4 |z-s|^2/{\rm Im}(z)$.  This proves
(\ref{ineq:abs3}).

By using (\ref{ineq:abs3}) with $s=z_\lambda$ we obtain that for $\lambda \in
R(E,\epsilon)$
$$
|z| \le 4 \mathrm{w}(z) + C\,,
$$
where $C$ depends only on $E$ and $\epsilon$.

Lemma \ref{Lx} extends Theorem \ref{T3.3} to all $x \in \mathbb{T}$ and  due to the previous inequality the statement of Theorem \ref{T3.2} follows. \qed

\section{Proofs of Lemma \ref{L3.1} and Lemma \ref{L3.2}}

In this section we will prove the bounds for $\mu_p$ stated in Lemma \ref{L3.1} and Lemma \ref{L3.2}. In order to do so we extend $\mu_p$, define some quantities to simplify the calculations and prove Proposition \ref{P3.1}. We prove Lemma \ref{L3.1} with the use of Proposition \ref{P3.1} and then prove Lemma \ref{L3.2}. 

Since in our lemmas we will use a compactification argument, we need to understand the behavior of $\mu_p(z_1,z_2,q_1\ldots q_M,\lambda)$ as $z_1$, $z_2$ approach the boundary of $\mathbb{H}$ and $\lambda$ approaches the real axis. Thus, it is natural to introduce the compactification $\overline{\mathbb{H}}^2 \times \mathbb{R}^M \times \overline{R}$. Here $ \overline{R}$ denotes the closure and $\overline{\mathbb{H}}$ is the compactification of $\mathbb{H}$ obtained by adjoining the boundary at infinity. (The word compactification is not quite accurate here because of the factor $\mathbb{R}$, but we will use the term nevertheless.)

The boundary at infinity is defined as follows.  We cover the upper half plane model of the hyperbolic plane $\mathbb{H}$ with the atlas $\mathcal{A}=\{(U_i, \psi_i)_{i=1,2}\}$. We have $U_1=\{z\in \mathbb{C} : {\rm Im}(z) > 0, |z| < C\}$, $\psi_1(z)=z$, $U_2 =\{z\in \mathbb{C} : {\rm
Im}(z) > 0, |z| > C\}$ and $\psi_2(z)=-1/z=u$. The boundary at infinity consists of the sets $\{{\rm Im}(z)=0\}$ and $\{{\rm Im}(u)=0\}$ in the respective charts. The compactification $\overline{\mathbb{H}}$ is the upper half plane with the boundary at infinity adjoined. We will use $i\infty$ to denote the point where $u=0$.

We defined $\mu_p$ for $z_1, z_2 \in \mathbb{H}^2$ and $\lambda \in R(E,\epsilon)$, and now we extend $\mu_p$ to an upper semi-continuous function on $\overline{\mathbb{H}}^2 \times \mathbb{R}^M \times {\overline R}$ by defining it as
\begin{eqnarray*}
  \mu_p(z_{1,0},z_{2,0},q_1\ldots q_M,\lambda_0) &=& \limsup_{z_1 \rightarrow z_{1,0}, z_2 \rightarrow z_{2,0}, \lambda\rightarrow \lambda_0}
  \mu_p(z_1,z_2,q_1\ldots q_M,\lambda)\ ,
\end{eqnarray*}
at points $(z_{1,0},z_{2,0})$ and $\lambda_0$ where it is not already defined.  Here, the points $(z_1,z_2)$ and $\lambda$ are approaching their limits in the topology of $\overline{\mathbb{H}}^2 \times \overline{R}$. For computational purposes we define the following quantities:
\begin{align*}
&A_n=(1+\lambda-q_1+z_1)(1+\lambda-q_2+\phi_1(z_1,q_1,\lambda))\ldots\\
&\hspace{7.8cm}(1+\lambda-q_n+\phi_{n-1}(z_1,q_1\ldots q_{n-1},\lambda))\\            
&A_m=(1+\lambda-q_{n+1}+z_1)(1+\lambda-q_{n+2}+\phi_1(z_1,q_{n+1},\lambda))\ldots \\
         & \hspace{6.9cm}       (1+\lambda-q_{M-1}+\phi_{m-1}(z_1,q_{n+1}\ldots q_{M-2},\lambda))\\
&C_n=(1+\lambda+z_\lambda)(1+\lambda+\phi_1(z_\lambda,0\ldots 0,\lambda))\ldots(1+\lambda+\phi_{n-1}(z_\lambda,0\ldots 0,\lambda))
\end{align*}
and similarly, if we replace $z_1$ with $z_2$ we obtain $B_n$ and $B_m$. $C_m$ is defined analogously to $C_n$ with $m$ factors in the product instead of $n$. If we expand the expressions for $A_n$, $B_n$ and $C_n$, respectively $A_m$, $B_m$ and $C_m$, defined above we can see that they are linear polynomials in the $z_i$ variable ($i$ can be $1$, $2$ or $\lambda$). It is also worth mentioning that $C_n \neq 0$, respectively $C_m \neq 0$, and if $\mathrm{Im}(z_i)>0$ then $A_n$, $A_m$, $B_n$, $B_m$ are also different from $0$. For more properties of these quantities see Section~\ref{appB}.

\noindent For the proof of Lemma \ref{L3.1} we need the following result:

\begin{prop} \la{P3.1} For all $z_1, z_2 \in \partial_{\infty}\overline{\mathbb{H}}^2$ and $\lambda\in E$,
\begin{eqnarray} 
\mu_0(z_1,z_2,0\ldots 0,\lambda) &< &1\,. \la{<1}
\end{eqnarray}
Here $E$ is any closed interval with $E \subset int(F \setminus S)$.
\end{prop}

\noindent {\it Remark}. In the case $m=n$, $\mu_0$ is symmetric in $z_1$ and $z_2$ and equals 1 at some points on the boundary. To then prove our desired result we would need to go back one more step in our recurrence formula and analyse a more complicated version of $\mu_p$.

\noindent {\it Proof of Proposition \ref{P3.1}}. Let us assume $n >m$. For $z_1, z_2 \in \mathbb{H}^2 \setminus (z_\lambda, z_\lambda)$ we write $z_1 = x_1+{\rm i}y_1$ and $z_2 = x_2+{\rm i}y_2$. Using these conventions, the triangle inequality and some simplifications we have
\begin{align}\la{3.7}
&\mathrm{w}(\phi(z_1, z_2,0\ldots 0, \lambda))= \frac{\left|(z_1-z_\lambda)(B_m C_m) + (z_2-z_\lambda)(A_n C_n)\right|^2}{(y_1|B_m|^2+y_2|A_n|^2)(|C_m|^2+|C_n|^2) {\rm Im}(z_\lambda)}\\
&\hspace{3.6cm}\leq \frac{(|z_1-z_\lambda||B_m C_m| + |z_2-z_\lambda||A_n C_n|)^2}{(y_1|B_m|^2+y_2|A_n|^2)(|C_m|^2+|C_n|^2) {\rm Im}(z_\lambda)} \nonumber
\end{align}
and a similar inequality for $\mathrm{w}(\phi(z_2, z_1,0\ldots 0, \lambda))$. These inequalities give us
\begin{eqnarray} \la{ND}
\mu_0(z_1,z_2,0\ldots 0,\lambda) &\leq& N/D\, 
\end{eqnarray} 
with
\begin{align} \la{Ne}
&N = \Big( (|z_1-z_\lambda||B_m C_m| + |z_2-z_\lambda||A_n C_n|)^2
(y_2|A_m|^2+y_1|B_n|^2)+&\\ \nonumber
&\hspace{4cm}(|z_2-z_\lambda| |A_mC_m|+|z_1-z_\lambda||B_nC_n|)^2
 (y_1|B_m|^2+y_2|A_n|^2)\Big) y_1 y_2& 
\end{align}
and
\begin{align}   \la{De}
 D= (|C_m|^2+|C_n|^2)(y_2|A_m|^2+y_1|B_n|^2) (y_1|B_m|^2+y_2|A_n|^2)&\\ \nonumber 
                         (|z_1-&z_\lambda|^2y_2+|z_2-z_\lambda|^2y_1)\,.
\end{align}
It is easy to check that $N/D \leq 1$ for $z_1, z_2 \in \mathbb{H}^2 \setminus (z_\lambda, z_\lambda)$, but we do not need this since the statement of our proposition only refers to the boundary \\$\dis \partial ( \overline{\mathbb{H}}^2 )  = \partial ( \overline{\mathbb{H}})  \times \partial(\overline{\mathbb{H}}) \cup \partial ( \overline{\mathbb{H}})  \times \mathbb{H} \cup\mathbb{H} \times \partial ( \overline{\mathbb{H}} ) $ where $\partial ( \overline{\mathbb{H}} ) =\mathbb{R} \cup \{{\rm i} \infty\}$. We know $\mu_0 \leq N/D \leq 1$, so we need to prove that at least one inequality is strict on the boundary. A few cases are to be considered:
\smallskip

{\it Case I:} Both $z_1$ and $z_2$ are on the real axis. Let $(z_{1,i},z_{2,i},\lambda_i) \rightarrow (z_1,z_2, \lambda)$ be a sequence that realizes the lim sup in the definition of $\mu_0$. Notice that since $y_{1,i} \rightarrow 0$ and $y_{2,i} \rightarrow 0$, $\lim\limits_{i \rightarrow \infty} N = \lim\limits_{i \rightarrow \infty} D = 0$ so the limit of $N/D$ may depend on the direction in which $z_{1,i}$ and $z_{2,i}$ approach $z_1$and $z_2$ . All the following variables will in fact be sequences determined by $(z_{1,i},z_{2,i},\lambda_i)$. We will sometimes suppress the index $i$ for simplicity. In order to deal with this undetermined case we use a blow-up, more precisely we write $y_1$ and $y_2$ in the following form:
\begin{eqnarray*}
  y_1&=& r_1 \omega_1 \\
  y_2 &=& r_1 \omega_2
\end{eqnarray*}
with $ \omega_1^2+ \omega_2^2 = 1$  and $r_1>0$, all functions of $z_1$ and $z_2$. By going to a subsequence if needed, assume $\omega_1$ and $\omega_2$ converge as $i \rightarrow \infty$. After cancelling a factor $r_1$, $N$ and $D$ in (\ref{ND}) become
\begin{align}\la{NI}
N= \Big( (|z_1-z_\lambda||B_m C_m| +|z_2-z_\lambda||A_n C_n|)^2
     (\omega_2|A_m|^2+\omega_1|B_n|^2)+ &\\(|z_2-z_\lambda||A_mC_m|+|z_1-
     z_\lambda||B_nC_n|)^2(\omega_1|B_m|^2+& \omega_2|A_n|^2)\Big)
     \omega_1 \omega_2 \,,\nonumber
\end{align}
\begin{align}\la{DI}
D= (|C_m|^2+|C_n|^2)(\omega_2|A_m|^2+
     \omega_1|B_n|^2)(\omega_1|B_m|^2+\omega_2|A_n|^2)&\\
     (|z_1-z_\lambda|^2&\omega_2+|z_2-z_\lambda|^2\omega_1)\,.\nonumber
\end{align}
Let us first look at the points on the boundary where  $D$ has a non vanishing limit. The points where $D \rightarrow 0$ will need extra blow-ups and will be analysed afterwards. 

We first show that $N/D \leq 1$ which is equivalent to proving the polynomial 
\begin{align*}
P(X,Y)=&X^2\omega_2\Big(\omega_1\omega_2|A_mB_mC_n|^2+\omega_1\omega_2|A_nB_nC_m|^2+\omega_2^2|A_nA_m|^2 (|C_m|^2+|C_n|^2)\Big)+\\
&Y^2\omega_1\Big(\omega_1\omega_2|A_mB_mC_n|^2+\omega_1\omega_2|A_nB_nC_m|^2+\omega_1^2|B_nB_m|^2 (|C_m|^2+|C_n|^2)\Big)-\\
&2XY\omega_1\omega_2|C_mC_n|\left(|A_nB_m|(\omega_2|A_m|^2+\omega_1|B_n|^2)+|A_mB_n|(\omega_1|B_m|^2+\omega_2|A_n|^2)\right)
\end{align*}
being positive; here $X=|z_1-z_\lambda|$ and $Y=|z_2-z_\lambda|$. It is easy to see that $P(X,Y)\geq 0$ since its discriminant has the form $ \left(|A_mB_m||C_n|^2-|A_nB_n||C_m|^2\right)^2\left(\omega_2|A_m|^2+\omega_1|B_n|^2\right)$\\$\left(\omega_1|B_m|^2+\omega_2|A_n|^2\right)$.

Let us now assume $ \mu_0 = 1$, so that $ \mu_0 = N/D=1$, and prove that the number of $\lambda$ values for which this can happen is finite. The condition $ \mu_0 = N/D$, which means equality in (\ref{3.7}), is equivalent to the existence of $p_1$, $p_2$, $s_1$, $s_2$ positive real numbers  and $\gamma$ and $\delta$ reals such that
\begin{align*}
&B_m C_m (z_1-z_\lambda) =  p_1 e^{\mathrm{i}\delta},\\
&A_n C_n (z_2-z_\lambda) =  p_2 e^{\mathrm{i}\delta},\\ 
&A_m C_m (z_2-z_\lambda) =  s_1 e^{\mathrm{i}\gamma},\\
&B_m C_m (z_1-z_\lambda) =  s_2 e^{\mathrm{i}\gamma},
\end{align*} 
which implies
\begin{align*}
&A_mB_m C_m^2 (z_1-z_\lambda) (z_2-z_\lambda)=  p_1 s_1e^{\mathrm{i}(\delta+\gamma)},\\
&A_n B_n C_n^2 (z_1-z_\lambda)(z_2-z_\lambda) =  p_2 s_2 e^{\mathrm{i}(\delta+\gamma)}, 
\end{align*} 
and therefore $\dis p_1 s_1 A_nB_n C_n^2 = p_2s_2 A_mB_m C_m^2$. This equality can be true iff $1).$ both sides are $0$ or $2).$ we have only non-zero terms which means, since $A_n$, $A_m$, $B_n$, $B_m$ are all real, $(C_n/C_m)^2$ must be real. Let us look at each of these two scenarios in detail.

$1).$  There are a few ways in which the right hand side of our equality can vanish.
\begin{itemize}
\item[a)] $p_1=0$; this implies $B_m=0$. Now, $N/D =1$ iff the discriminant mentioned above is $0$ which can happen if:
\begin{itemize}
\item $|A_n|=0$ which means we are in the case $D=0$ discussed later;
\item $\omega_2 =0$,  we are again in the case $D=0$,
\item $\omega_1=0$, $|A_m|=0$  we are in the case $D=0$,
\item $\omega_2=0$, $|B_n|=0$  we are in the case $D=0$,
\item $|B_n|=0$; in this case $B_m=B_n=0$ and according to Lemma \ref{L3.4} this can happen for at most a finite number of $\lambda$ values which will be included in $S$.
\end{itemize}
\item[b)] $s_1=0$; this implies $A_m=0$ and the analysis will be almost identical to the one in  a).
\item[c)] $A_n=0$; this implies $p_2=0$ and we are in a similar case to a).
\item[d)] $B_n=0$; this implies $s_2=0$ and we are in a similar case to b).
\end{itemize}
We should also notice $C_n \neq 0$ and $C_m \neq 0$. 

$2).$ $\dis \left(\frac{C_n}{C_m}\right)^2 \in \mathbb{R}$. We have two possibilities:
\begin{itemize}
\item $\dis \frac{C_n}{C_m} \in \mathbb{R}$ which according to Lemma \ref{L3.4} can be true for at most a finite number of $\lambda$ values which will be included in $S$,
\item $\dis \frac{C_n}{C_m} = r\, \mathrm{i}$, $r \in \mathbb{R}$, which according to Lemma \ref{L3.4} can be true for at most a finite number of $\lambda$ values which will be included in $S$.
\end{itemize}

The points where $D \rightarrow 0$ have to be analysed separately. There are a few ways in which our denominator can vanish. The first and the last term in the expression for $D$ cannot be zero, it is only the two middle factors that can become $0$. The following situations arise: 

\noindent {\it Scenario 1:} $(\omega_2|A_m|^2+\omega_1|B_n|^2) \nrightarrow 0$ and $(\omega_1|B_m|^2+\omega_2|A_n|^2) \rightarrow 0$. This situation can happen if:
\begin{itemize}
\item $\omega_2 \rightarrow 0$ and $|B_m|^2 \rightarrow 0$, or
\item $|A_n|^2 \rightarrow 0$ and  $\omega_1 \rightarrow 0$. Since the analysis of these two cases is almost identical, we will only look at this second one. We need to consider a blow-up:
\begin{eqnarray*}
 |A_n|^2 &=& r_2 \sin(\alpha)\\
 \omega_1&=& r_2 \cos(\alpha) \,
\end{eqnarray*}
with $r_2 >0$ and $\alpha \in [0, \pi/2]$ functions of $z_1, z_2$ and $\lambda$. With this new blow-up we have
\begin{align*}
&N= \Big( (|z_1-z_\lambda||B_m C_m| +|z_2-z_\lambda|(r_2 \sin(\alpha))^{1/2} |C_n|)^2
     (\omega_2|A_m|^2+ r_2 \cos(\alpha)|B_n|^2)\\
&+(|z_2-z_\lambda||A_mC_m|+|z_1-z_\lambda||B_nC_n|)^2( r_2 \cos(\alpha)|B_m|^2+ \omega_2 r_2 \sin(\alpha))\Big) r_2 \sin(\alpha) \omega_2 
\end{align*}
and
\begin{align*}
&D= (|C_m|^2+|C_n|^2)(\omega_2|A_m|^2+ r_2 \cos(\alpha)|B_n|^2)(  r_2 \cos(\alpha) |B_m|^2+\omega_2 r_2 \sin(\alpha))\\
&\cdot (|z_1-z_\lambda|^2 \omega_2+|z_2-z_\lambda|^2  r_2 \cos(\alpha))\,.
\end{align*} 
By going to a subsequence if needed we can assume that $r_{2,i}$, $B_{m,i}$, $C_{m,i}$, $C_{n,i}$, $\omega_{2,i}$, $\alpha_{i}$ converge to $0$, $\overline{B}_m$, $\overline{C}_m$, $\overline{C}_n$, $1$, $\overline{\alpha}$ respectively (recall that $B_{m,i}$ is a linear polynomial in $z_{2,i}$).  In this situation we find
$$
\mu_0 \leq \frac{| \overline{B}_m|^2
|\overline{C}_m|^2\cos(\overline{\alpha})}{(| \overline{C}_m|^2+| \overline{C}_n|^2)(\sin( \overline{\alpha})+
 |\overline{B}_m|^2\cos(\overline{\alpha}))}<1\,.
$$ 
\item The last case under this scenario is $|A_n|^2 \rightarrow 0$ and $|B_m|^2 \rightarrow 0$. After a blow-up of the form
\begin{eqnarray*}
 |A_n|^2 &=& r_3 \cos(\beta)\\
 |B_m|^2&=& r_3 \sin(\beta) \,
\end{eqnarray*}
with $r_3 >0$ and $\beta \in [0, \pi/2]$ functions of $z_1, z_2$ and $\lambda$, we have
\begin{align*}
N= \Big( (|z_1-z_\lambda| (\sin(\beta))^{1/2}|C_m| +|z_2-z_\lambda|( \cos(\beta))^{1/2}|C_n|)^2
     (\omega_2|A_m|^2+\omega_1|B_n|^2)+ &\\(|z_2-z_\lambda||A_mC_m|+|z_1-
     z_\lambda||B_nC_n|)^2(\omega_1 \sin(\beta)+ \omega_2 \cos(\beta))\Big)
     \omega_1 \omega_2 \,,
\end{align*}
\begin{align*}
D= (|C_m|^2+|C_n|^2)(\omega_2|A_m|^2+
     \omega_1|B_n|^2)(\omega_1 \sin(\beta)+\omega_2 \cos(\beta))&\\
     (|z_1-z_\lambda|^2\omega_2+&|z_2-z_\lambda|^2\omega_1)\,.
\end{align*}
Now, the new expression for $D$ would vanish only if $\omega_1=0$ and $\cos(\beta)=0$, or $\sin(\beta)=0$ and $\omega_2=0$ respectively. This means we have the cases $\omega_1=0$ and $|A_n|^2=0$, or $|B_m|^2=0$ and $\omega_2=0$ which were already discussed. Otherwise, the expression $N/D$ is well defined and by arguments similar to the ones before strictly less than unity. 
\end{itemize}

\noindent {\it Scenario 2:} $(\omega_2|A_m|^2+\omega_1|B_n|^2) \rightarrow 0$ and $(\omega_1|B_m|^2+\omega_2|A_n|^2) \nrightarrow 0$. This can happen if:
\begin{itemize}
\item $\omega_2 \rightarrow 0$ and $|B_n|^2 \rightarrow 0$, or
\item $|A_m|^2 \rightarrow 0$ and  $\omega_1 \rightarrow 0$ or 
\item $|A_m|^2 \rightarrow 0$ and $|B_n|^2 \rightarrow 0$.
\end{itemize}
Since this scenario is very much the same as the previous one we will not discuss it any further.  

\noindent {\it Scenario 3:} $(\omega_2|A_m|^2+\omega_1|B_n|^2) \rightarrow 0$ and $(\omega_1|B_m|^2+\omega_2|A_n|^2) \rightarrow 0$. This situation can happen if:
\begin{itemize}
\item $\omega_2 \rightarrow 0$, $|B_n|^2 \rightarrow 0$ and $|B_m|^2 \rightarrow 0$, or
\item $\omega_1 \rightarrow 0$, $|A_n|^2 \rightarrow 0$ and $|A_m|^2 \rightarrow 0$. Again, due to the symmetry of our expression, it is enough to look at this second case. We need a blow-up of the  form \begin{eqnarray*}
  \omega_1&=& r_3\, \gamma_1 \\
  |A_n|^2 &=& r_3 \,\gamma_2 \\
  |A_m|^2 &=& r_3 \,\gamma_3
\end{eqnarray*}
with $ \gamma_1^2+ \gamma_2^2+ \gamma_3^2 = 1$  and $r_3>0$, functions of $z_1, z_2$ and $\lambda$.
\begin{align*}
N= \Big( (|z_1-z_\lambda||B_m C_m| +|z_2-z_\lambda|(r_3\gamma_2)^{1/2} |C_n|)^2
     (\omega_2\gamma_3+\gamma_1|B_n|^2)+ &\\(|z_2-z_\lambda|(r_3\gamma_2)^{1/2}|C_m|+|z_1-
     z_\lambda||B_nC_n|)^2(\gamma_1|B_m|^2+& \omega_2 \gamma_2)\Big)
     \gamma_1 \omega_2 \,,
\end{align*}
\begin{align*}
D= (|C_m|^2+|C_n|^2)(\omega_2\gamma_3+
     \gamma_1|B_n|^2)(\gamma_1|B_m|^2+\omega_2 \gamma_2)&\\
     (|z_1-z_\lambda|^2&\omega_2+|z_2-z_\lambda|^2 (r_3\gamma_2)^{1/2})\,.
\end{align*}
By going to a subsequence if needed we can assume that $r_{3,i}$, $B_{m,i}$, $C_{m,i}$, $C_{n,i}$, $\omega_{2,i}$, $\gamma_{1,i}$, $\gamma_{2,i}$, $\gamma_{3,i}$ converge to $0$, $\overline{B}_m$, $\overline{C}_m$, $\overline{C}_n$, $1$, $\overline{\gamma_1}$, $\overline{\gamma_2}$, $\overline{\gamma_3}$ respectively. In this situation we obtain
$$
\mu_0 \leq \frac{\Big( |\overline{B}_m \overline{C}_m|^2 (\overline{\gamma_3}+\overline{\gamma_1}|\overline{B}_n|^2)+|\overline{B}_n\overline{C}_n|^2(\overline{\gamma_1}|\overline{B}_m|^2+ \overline{\gamma_2})\Big) \overline{\gamma_1}} {(|\overline{C}_m|^2+|\overline{C}_n|^2)(\overline{\gamma_3}+
     \overline{\gamma_1}|\overline{B}_n|^2)(\overline{\gamma_1}|\overline{B}_m|^2+ \overline{\gamma_2})} <1\,,
$$
provided the denominator does not vanish. If $\overline{\gamma}_1=\overline{\gamma}_2=0$, $\overline{\gamma}_1=\overline{\gamma}_3=0$, $\overline{\gamma}_3=|\overline{B}_n|^2=0$ or/and $|\overline{B}_m|^2=\overline{\gamma}_2=0$ extra blow-ups are needed, but the limiting value for $N/D$ stays strictly less than $1$. As an example, if we consider the extra blow-up given by  $\overline{\gamma_3}= r_4 \gamma_3$, $\overline{\gamma_1}=r_4 \gamma_1$, $\gamma_1^2+\gamma_3^2=1$  and $r_4>0$ we have
$$
\mu_0 \leq \frac{|\overline{B}_n\overline{C}_n|^2 \gamma_1} {(|\overline{C}_m|^2+|\overline{C}_n|^2)(\gamma_3+
     \gamma_1|\overline{B}_n|^2)} <1\,.
$$
\item $|A_m|^2 \rightarrow 0$, $|B_n|^2 \rightarrow 0$, $|B_m|^2 \rightarrow 0$ and $|A_m|^2 \rightarrow 0$. After a needed blow-up the expressions will look similar to the ones in (\ref{NI}) and (\ref{DI}), but in the blown-up variables. 
\end{itemize}
\smallskip

{\it Case II:} Both $z_1$ and $z_2$ are ${\rm i}\infty$. Let $(z_{1,j},z_{2,j}, \lambda_j) \in \overline{\mathbb{H}^2} \times \overline{R}$ be a sequence that realizes the lim sup in the definition of $\mu_0$. We sometimes suppress the index $j$ for simplicity.  We consider the change of 
variables, $u_1=-\frac{1}{z_1}$, $u_2=-\frac{1}{z_2}$ and $u_\lambda=-\frac{1}{z_\lambda}$; now, both $u_1$ and $u_2$ approach $0$. With these new variables, $N$ and $D$ from (\ref{ND}) are given by 
\begin{align*}
&N= \Big( \big(| u_\lambda-u_1 ||u_2 u_\lambda B_m C_m| +| u_\lambda-u_2 ||u_1 u_\lambda
     A_n C_n|\big)^2\big( {\rm Im}(u_2)|u_1A_m|^2+{\rm Im}(u_1)|u_2B_n|^2\big)\\
     &\hspace{6cm} +\big(| u_\lambda-u_2||u_1u_\lambda A_mC_m|+ | u_\lambda-u_1||u_2  u_\lambda B_nC_n|\big)^2 \\
&\hspace{5.9cm}\cdot \big({\rm Im}(u_1)|u_2 B_m|^2+{\rm Im}(u_2)|u_1 A_n|^2\big)\Big){\rm Im}(u_1)  {\rm Im}(u_2),
\end{align*}
\begin{align*}
&D= \left(|C_m|^2+|C_n|^2\right) |u_\lambda|^2 \left( {\rm Im}(u_2)|u_1 A_m|^2+{\rm Im}(u_1)|u_2 B_n|^2 \right)\\
&\hspace{1.5cm}\cdot \left({\rm Im}(u_1)|u_2 B_m|^2+ {\rm Im}(u_2)|u_1 A_n|^2 \right) 
 \left(| u_\lambda-u_1|^2  {\rm Im}(u_2)+| u_\lambda-u_2|^2{\rm Im}(u_1)\right).
\end{align*}
Since both sequences $u_{1,j}$ and $u_{2,j}$ are approaching $0$, we can write \\$\dis u_{1,j}=r_j \cos(\gamma_j){\rm e}^{{\rm i} x_j}$ and $u_{2,j}=r_j \sin(\gamma_j){\rm
e}^{{\rm i} y_j}$, with $\gamma_j, x_j, y_j \in[0,\pi/2]$.  By going to a subsequence if needed and recalling that $A_{n,j}$, $A_{m,j}$, $B_{n,j}$ and $B_{m,j}$ are linear polynomial is $z_1$ and $z_2$, we can assume that $r_{j}$, $u_{1,j}A_{n,j}$, $u_{1,j}A_{m,j}$, $u_{2,j}B_{n,j}$, $u_{2,j}B_{m,j}$, $u_{\lambda,j}C_{n,j}$, $u_{\lambda,j}C_{m,j}$, $\gamma_{j}$, $x_{j}$, $y_{j}$ converge to $0$, $\overline{A}_n$, $\overline{A}_m$, $\overline{B}_n$, $\overline{B}_m$, $\overline{C}_n$, $\overline{C}_m$, $\overline{\gamma}$, $\overline{x}$, $\overline{y}$ respectively. After cancelling the common factor of $r_{j}$ and $|u_{\lambda,j}|$ in the above expressions for $N$ and $D$ and taking the limit we get
\begin{align*}
&N= \Big( \big(|\overline{B}_m \overline{C}_m| + |\overline{A}_n \overline{C}_n|\big)^2\big(\sin(\gamma)\sin(y)|\overline{A}_m|^2+ \cos(\gamma)\sin(x)|\overline{B}_n|^2\big)\\
&\hspace{3cm}+\big(|\overline{A}_m\overline{C}_m|+|\overline{B}_n\overline{C}_n|\big)^2 \big(\sin(\gamma)\sin(y)|\overline{A}_n|^2+\cos(\gamma)\sin(x)|\overline{B}_m|^2\big)\Big)\\
&\hspace{8.5cm}\cdot  \sin(x)\sin(y)\sin(\gamma)\cos(\gamma)\,,
\end{align*}
\begin{align*}
&D=\big(|\overline{C}_m|^2+|\overline{C}_n|^2\big)\big(\sin(\gamma)\sin(y)|\overline{A}_m|^2+ \cos(\gamma)\sin(x)|\overline{B}_n|^2\big)\\
&\hspace{1cm}\big(\sin(\gamma)\sin(y)|\overline{A}_n|^2+\cos(\gamma \sin(x)|\overline{B}_m|^2\big)\big(\sin(x)\cos(\gamma)+\sin(y)\sin(\gamma)\big)\,.
\end{align*}
If we compare this with {\it Case I} and consider $|z_1 - z_\lambda|=|z_2-z_\lambda|$, $y_1= \cos(\gamma)\sin(x)$ and $y_2=\sin(\gamma)\sin(y)$ we can see that we are in a similar situation to the one in {\it Case I} . 
\smallskip

 {\it Case III:} $z_1 \in \mathbb{R}$ and $z_2 ={\rm i} \infty$, respectively $z_2 \in \mathbb{R}$ and $z_1 ={\rm i} \infty$. We consider again a sequence that realizes the lim sup in the definition of $\mu_0$ and we use the same change of variables for $z_2$, as before. Since $z_1 \rightarrow \mathbb{R}$ and $u_2 \rightarrow 0$  we can write $u_2=r e^{i y_2}$ with $r>0$ and ${\rm Im}(z_1)=y_1$. After we cancel in both $N$ and $D$ a factor of $r^6$ and $|u_\lambda|^2$ we have 
\begin{align} \la{n3}
&N= \Big( \big(| z_1-z_\lambda | |u_2 B_m u_\lambda C_m| +| r e^{i y_2} -u_\lambda||A_n C_n|\big)^2\big(r \sin(y_2)|A_m|^2+y_1|u_2 B_n|^2\big) \\  
 &+\big(|r e^{i y_2}-u_\lambda||A_m u_\lambda C_m|+|z_1- z_\lambda| |u_2 B_n u_\lambda C_n|\big)^2\big(y_1|u_2  B_m|^2+r \sin(y_2)| A_n|^2\big)\Big)r y_1 \sin(y_2), \nonumber 
\end{align}
\begin{align} \la{d3}
&D= (|u_\lambda C_m|^2+|u_\lambda C_n|^2)\left(r \sin(y_2)|A_m|^2+y_1|u_2 B_n|^2 \right)\left(y_1|u_2 B_m|^2+r \sin(y_2)|A_n|^2 \right)\nonumber\\ 
&\hspace{5cm}     \left(| z_1-z_\lambda|^2  r \sin(y_2)|u_\lambda|^2+|r e^{i y_2} - u_\lambda|^2y_1\right).
\end{align}
If we compare it with {\it Case I} and consider $|z_1 - z_\lambda|=|(z_1-z_\lambda)u_\lambda|$, $|z_2-z_\lambda|=|r e^{i y_2}-u_\lambda|$, $y_1= y_1$, $y_2=r \sin(y_2)$, $|A_n|=|A_n|$, $|A_m|=|A_m|$, $|B_n|=|u_2 B_n|$ and $|B_m|=|u_2 B_m|$ we can see that the blow-ups needed are similar to the ones in {\it Case I} and we can conclude $N/D <1$. 
\smallskip

{\it Case IV:} $z_1 \in \mathbb{H}$ and $z_2 \in \mathbb{R}$, respectively $z_2 \in \mathbb{H}$ and $z_1 \in \mathbb{R}$.  As before, we take a sequence that realizes the lim sup in the definition of $\mu_0$. If we look at the expressions for $D$ given by (\ref{De}) we can see that we can have the following three undetermined cases: $y_2 \rightarrow 0$ and $|B_n|^2 \rightarrow 0$, similarly $y_2 \rightarrow 0$ and $|B_m|^2 \rightarrow 0$, or $y_2 \rightarrow 0$, $|B_n|^2 \rightarrow 0$ and $|B_m|^2 \rightarrow 0$. The analysis of these blow-up cases can be done in a similar manner with the one from {\it Case I} and we can conclude that $N/D$ is strictly less than $1$.
\smallskip

{\it Case V:} $z_1 \in \mathbb{H}$ and $z_2 = {\rm i} \infty$, respectively $z_2 \in \mathbb{H}$ and $z_1 = {\rm i} \infty$. We take a sequence that realizes the lim sup and we consider the same change of variables as in {\it Case III}. With the same notations $u_2=r e^{i y_2}$ with $r>0$ and ${\rm Im}(u_1)=y_1$ we obtain the same expressions for $N$ and $D$ as in (\ref{n3}) and (\ref{d3}). With similar blow-ups with the ones in {\it Case III} we can conclude that also in this last case the limiting value for $N/D$ is strictly less than $1$. \qed

\vspace{5mm}

\noindent {\bf Proof of Lemma \ref{L3.1}.} To prove the lemma it is enough to show that
\begin{eqnarray*}
  \mu_p(Z, Q, \lambda) &<& 1
\end{eqnarray*}
for $(Z, Q, \lambda)$ in the compact set $\partial_\infty (\overline{\mathbb{H}}^2)
\times \{0\}^M \times E$, since this implies that for some $\epsilon
> 0$, the upper semi-continuous function $\mu_p(Z, Q, \lambda)$
is bounded by $1-2\epsilon$ on the set, and by $1-\epsilon$ in some
neighborhood.

Let us rewrite $\mu_p$ in terms of $\mu_0$.
\begin{align*}
\mu_p(Z,Q, \lambda) = \frac{\mathrm{w}^{1+p}(\phi(z_1,z_2,q_1\ldots q_M,\lambda))+\mathrm{w}^{1+p}(\phi(z_2,z_1,q_1\ldots q_M,\lambda))}{ \mathrm{w}^{1+p}(z_1)+ \mathrm{w}^{1+p}(z_2)} &\\
\leq \left( \frac{\mathrm{w}(\phi(z_1,z_2,q_1\ldots q_M,\lambda))+\mathrm{w}(\phi(z_2,z_1,q_1\ldots q_M,\lambda))}{ \mathrm{w}(z_1) + \mathrm{w}(z_2)}\right)^{1+p} &\cdot \\ \cdot \frac{1}{\nu_1^{1+p}+\nu_2^{1+p}}\,,&
\end{align*}
where 
$$
\nu_i=\frac{\mathrm{w}(z_i)}{\mathrm{w}(z_1)+\mathrm{w}(z_2)}\,,\,\,\rm{for}\,\,\,i=1,\,2\,.
$$
Since we are concentrating on the boundary of $\overline{\mathbb{H}}^2$, we need the following blow-up
\begin{eqnarray*}
  \chi(z_1) = \frac{1}{\mathrm{w}(z_1)} = R_1 \Omega_1 \\
  \chi(z_2) = \frac{1}{\mathrm{w}(z_2)} = R_1 \Omega_2
\end{eqnarray*}
where $R_1, \Omega_1$ and $\Omega_2$ are defined
as functions of $z_1$ and $z_2$ with the property $\Omega_1^2+\Omega_2^2=1$. Using the result in Proposition \ref{P3.1} we have
\begin{eqnarray*}
\mu_p |_{\partial_\infty (\overline{\mathbb{H}})^2 \times \{0\}^M \times E} < \left(\mu_0 |_{\partial_\infty(\overline{\mathbb{H}}^2) \times \{0\}^M \times E}\right)^{1+p} \frac{\left(\Omega_1+\Omega_2 \right)^{1+p}}{\Omega_1^{1+p}+\Omega_2^{1+p}} 
\leq (1-\epsilon) 2^p < 1\,,
\end{eqnarray*}
for sufficiently small $p$.\qed

\vspace{5mm}

\noindent {\bf Proof of Lemma \ref{L3.2}.} Each term in the sum appearing in $\mu_p$
can be estimated

\begin{align*}
\frac{\mathrm{w}
^{1+p}(\phi(z_1,z_2,q_1\ldots q_M,\lambda))}{\mathrm{w}
^{1+p}(z_1) + \mathrm{w} ^{1+p}(z_2)} &= \frac{(\mathrm{w}(z_1)  + \mathrm{w}(z_2))^{1+p}}{\mathrm{w}^{1+p}(z_1) + \mathrm{w}^{1+p}(z_2)} \left(
\frac{\mathrm{w}(\phi(z_1,z_2,q_1\ldots q_M,\lambda))}{\mathrm{w}(z_1) + \mathrm{w}(z_2)}\right)^{1+p} \\ &\le 2^p \left(\frac{\mathrm{w}(\phi(z_1,z_2,q_1\ldots q_M,\lambda))}{\mathrm{w}(z_1)  +
\mathrm{w}(z_2)}\right)^{1+p} .
\end{align*}
Now it is enough to prove that 
$\dis 
\frac{\mathrm{w}(\phi(z_1,z_2,q_1\ldots q_M,\lambda))}{\mathrm{w}(z_1) + \mathrm{w}(z_2)} \le C\prod_{i=1}^M (1+|q_i|^2)\,,
$
since this bounds each term in $\mu_p$ by the desired quantity. With the notations introduced at the beginning of this section, $\dis \phi_n(z_1,q_1\ldots q_n,\lambda) = -\frac{A_{n-1}}{A_n}$ and applying Cauchy-Schwarz inequality twice we get
\begin{align*}
&\frac{ \mathrm{w}(\phi(z_1,z_2,q_1\ldots q_M,\lambda))}{ \mathrm{w}(z_1) + \mathrm{w}(z_2)}\\
&=\frac{\left|1+z_{\lambda}\phi_n(z_1,q_1\ldots q_n,\lambda)+
    z_{\lambda} \phi_m(z_2,q_{n+1} \ldots q_{n+m},\lambda)+z_{\lambda}
    (\lambda-q_M) \right|^2}{{\rm Im}(\phi_n(z_1,q_1 \ldots q_n,
    \lambda))+{\rm Im}(\phi_m(z_2,q_{n+1} \ldots q_{n+m},\lambda))+{\rm
    Im}(\lambda)} \cdot \frac{1}{\sum\limits_{i=1}^{2}\frac{|z_i-
    z_\lambda|^2}{{\rm Im}(z_i)}} \\
& \leq \frac{\left|A_n B_m - z_{\lambda}A_{n-1}B_m- z_{\lambda} B_{m-1}A_n+A_n
    B_m z_{\lambda}(\lambda-q_M) \right|^2}{{\rm Im}(z_1)
    |B_m|^2+{\rm Im}(z_2)|A_n|^2} \cdot \frac{1}{\sum\limits_{i=1}^{2}\frac{|z_i-
    z_\lambda|^2}{{\rm Im}(z_i)}}\\   
&\leq \big(1+|z_\lambda|^2\big) \Big(\frac{|A_nB_m|^2}{{\rm Im}(z_1)|B_m|^2+
    {\rm Im}(z_2)|A_n|^2}+\big(|A_nB_m|^2+|A_{n-1}B_m+A_nB_{m-1}|^2\big)\cdot\\
    &\hspace{7cm} \frac{\big(1+|q_{M}-\lambda|^2\big)}{{\rm Im}(z_1)|B_m|^2+
    {\rm Im}(z_2)|A_n|^2}\Big)\cdot \frac{1}{\sum\limits_{i=1}^{2} 
    \frac{|z_i-z_\lambda|^2}{{\rm Im}(z_i)}} \\
 &\leq (1+|z_\lambda|^2) \left(\frac{|A_n|^2(2+|q_{M}-\lambda|^2)}{{\rm
    Im}(z_1)}+2(1+|q_{M}-\lambda|^2)\left(
    \frac{|A_{n-1}|^2}{{\rm Im}(z_1)}+\frac{|B_{m-1}|^2}{
    {\rm Im}(z_2)}\right)\right) \cdot\frac{1}{\sum\limits_{i=1}^{2}
    \frac{|z_i-z_\lambda|^2}{{\rm Im}(z_i)}} .
 \end{align*}
Since $\phi_n$ is a fractional linear transformation with coefficients given by the product matrix 
$ \prod\limits_{ i=1}^n 
\begin{pmatrix}
0 & -1 \\
1 & 1+\lambda-q_i 
\end{pmatrix}$, $A_n$, the denominator of $\phi_n$, is a linear polynomial in $z$ whose coefficients can be bounded above. We get
$\,\,\,\,\,\,\dis |A_n| \leq n\, |1+z_1|\, \prod\limits_{\substack{|1+\lambda-q_i| \geq 1\\ i=1}}^n |1+\lambda-q_i|$ which implies, $\dis |A_n|^2 \leq C \left(1+|z_1|^2\right) \prod\limits_{i=1}^n \left(1+|q_i|^2\right)$.

\intremark{ We have $\phi_i$ of the form
\begin{eqnarray*}
\phi_1 &=& \frac{-1}{z+a_1} \\
\phi_2&=& \frac{-(z+a_1)}{a_2 z+ a_1a_2-1} \\
\phi_3&=& \frac{-(a_2z+a_1a_2-1)}{(a_2a_3-1)z+a_1a_2a_3 -a_3 - a_1}\\
\ldots\,.
\end{eqnarray*}
Since $A_n$ is the denominator of $\phi_n$, it is a linear polynomial in $z$ where each of its coefficients can be bounded above. We get
\begin{eqnarray*}
|A_n| &\leq& n\, |1+z|\, \prod\limits_{\substack{|a_i| \geq 1\\ i=1}}^n |a_i|\,.
\end{eqnarray*}
Therefore,
\begin{eqnarray*}
|A_n|^2 &\leq& C \prod\limits_{i=1}^n \left(1+|q_i|^2\right)\left(1+|z|^2\right)
\end{eqnarray*}
}   
Going back to our inequality we have 
\begin{align*}
\frac{ \mathrm{w}(\phi(z_1,z_2,q_1\ldots q_{M},\lambda))}{ \mathrm{w}(z_1) + \mathrm{w}(z_2)}
    &\leq C \Big(  C_1 (1+|q_{M}|^2)\prod\limits_{i=1}^n (1+|q_i|^2)
    \frac{1+|z_1|^2}{{\rm Im}(z_1)}\\
    &+C_2 (1+|q_{M}|^2) \prod\limits_{i=1}^{n-1} (1+|q_i|^2)
    \frac{1+|z_1|^2}{{\rm Im}(z_1)}\\
    &+C_3(1+|q_{M}|^2)\prod\limits_{k=n+1}^{n+m-1} (1+|q_i|^2)
    \frac{1+|z_2|^2}{{\rm Im}(z_2)} \Big)
    \cdot \frac{1}{\sum\limits_{i=1}^{2}
    \frac{|z_i-z_\lambda|^2}{{\rm Im}(z_i)}}\,.
\end{align*}
Choose the compact set $\cal K$ such that $\sum\limits_{i=1}^{2}
|z_i-z_\lambda|^2 / {\rm Im}(z_i)\ge C >0$ for some constant $C$ and
$(z_1,z_2)\in {\cal K}^c$. Then we can estimate each term depending
on whether $z_i$ is close to $z_\lambda$. If $z_j$, $j=1,2$, is sufficiently close, then ${\rm Im}(z_j)$ is bounded below and
$|z_j|$ is bounded above by a constant. Thus
$$
{\rm Im}(z_j)\sum_{i=1}^{2} |z_i-z_\lambda|^2 / {\rm Im}(z_i)\ge
{\rm Im}(z_j)C \ge C'> 0
$$
and $ 1+|z_j|^2 \le C \,, $ so we are done. If $z_j$, $j=1,2$, is far
from $z_\lambda$,
$$
{\rm Im}(z_j)\sum_{i=1}^{2} |z_i-z_\lambda|^2 / {\rm Im}(z_i) \ge
|z_j-z_\lambda|^2 \ge C(1+|z_j|^2)
$$
so $1+|z_j|^2 / \left({\rm Im}(z_j)\sum\limits_{i=1}^{2}
|z_i-z_\lambda|^2/ {\rm Im}(z_i)\right) \le C$ again. \qed

\newpage

\section{Additional Results} \la{appA}

This section contains two theorems on absolutely continuous spectrum. The first one gives a sufficient condition for a measure to be absolutely continuous with respect to the Lebesgue measure on an interval and the latter gives a sufficient condition for a random Schr{\"o}dinger operator to exhibit purely absolutely continuous spectrum on some interval.

\subsection{A Criterion for Absolutely Continuous Spectrum}

Let $\mu$ be a finite measure on $\mathbb{R}$; its Stieltjes (or Borel)
transform  $F$ is given by
\begin{equation*}
F(z)=\int \frac{d\mu(t)}{t-z}
\end{equation*}
for $z=x+ i \,y$ with $y>0$.  The following criterion has been proven in \cite{S-2} for $\limsup$ and we reproduce it here for $\liminf$.

\begin{prop} \la{Simon} Let $(a,b)$ be a finite interval and let $p>0$. Suppose
\begin{equation*}
\liminf_{y \rightarrow 0} \int_a^b |F(x+{\rm i} y)|^{1+p} dx < \infty\,.
\end{equation*}
Then $\mu$ is absolutely continuous with respect to the Lebesgue measure on $(a,b)$. 
\end{prop}

\proof  Since $\liminf \limits_{y \rightarrow 0} \int_a^b
|F(x+{\rm i} y)|^{1+p} dx < \infty$, there exists a sequence $y_n \rightarrow 0$ such
that $\sup\limits_{n} \int_a^b |F(x+{\rm i}
y_n)|^{1+p} dx < C$, where $C$ is some constant. Define
$d\mu_{y_n}(x) = \pi^{-1} {\rm Im}( F(x+{\rm i}y_n))dx$. Then by~\cite{S-1}, $d\mu_{y_n} \rightarrow d\mu$ weakly, as $n \rightarrow
\infty$. That is, for $f$ a continuous function of compact support
we have $\lim \limits_{n \rightarrow \infty} \int f(x)d\mu_{y_n}(x)
= \int f(x) d\mu(x)$. Let $f$ be a continuous function supported on
$(a,b)$, then
\begin{eqnarray*}
\left|\int_a^b f(x) d\mu(x)\right| &=& \lim \limits_{n \rightarrow
\infty} \left|\int_a^b
f(x)d\mu_{y_n}(x)\right| \\
&=& \lim \limits_{n \rightarrow \infty} \pi^{-1} \left|\int_a^b f(x)
{\rm Im}( F(x+{\rm
i}y_n))dx \right| \\
&\leq& \lim \limits_{n \rightarrow \infty} \left(||f||_{1+1/p}
||{\rm Im}
(F(x+{\rm i}y_n)||_{1+p}\right) \\
&\leq& C ||f||_{1+1/p}\,.
\end{eqnarray*}
This implies that $d\mu(x) = g(x) dx$ for some $g \in L^{1+p}$. \qed

\subsection{Bounds on the Green Function at an Arbitrary Site}

The following lemma proves that assuming we have a bound for the forward Green functions $G^x(\lambda)$ for all $x \in \mathbb{T}_p$, we can obtain a bound for all the diagonal matrix elements $G_x(\lambda)$, $x \in \mathbb{T}$, of the Green function. 

\begin{lem} \la{Lx} Let $F$ be the open interior of the absolutely continuous spectrum of $\Delta$. Suppose that for any $x \in \mathbb{T}_p$
$$
\sup_{\lambda \in R(E,\epsilon)} \mathbb{E} \left(  \mathrm{w}^{1+p}\left(\left\langle {\delta_x,(H^x -\lambda)^{-1}\delta_x} \right\rangle\right) \right) < \infty \; ,
$$
for some closed subinterval $E \subset F$, $\epsilon>0$ and $0<p<1$.
Then, for every $x \in \mathbb{T}$, we also have
$$
\sup_{\lambda \in R(E,\epsilon)} \mathbb{E} \left( \mathrm{w}^{1+p}\left(\left\langle {\delta_x,(H -\lambda)^{-1}\delta_x} \right\rangle\right) \right) < \infty \; .
$$
\end{lem}

\proof Suppose we pick an arbitrary node $x_0$ in $\mathbb{T}$ and we consider its corresponding diagonal matrix element of the Green function for the whole tree $\mathbb{T}$, $G_{x_0}(\lambda)=\left\langle {\delta_{x_0},(H -\lambda)^{-1}\delta_{x_0}}\right\rangle$. We rearrange the nodes, if needed, such that $x_0$ becomes the origin of the tree. For this origin, we have $G_{x_{0}}(\lambda) = G^{x_{0}}(\lambda)$. Looking at the vertices in the future of $x_0$, we can see that after a finite number of steps, on each branch,the future tree will be a copy of the original tree. Let us denote by  $x_i$ the nodes where such a copy starts. An example of such a rearrangement is illustrated in the picture below. 
\newline We know from the hypothesis that 
$$
\sup_{\lambda \in R(E,\epsilon)} \mathbb{E} \left( \mathrm{w}^{1+p}\left( \left\langle \delta_{x_i},(H^{x_i} -\lambda)^{-1}\delta_{x_i} \right\rangle\right) \right) < \infty \; .
$$
Starting with these nodes and using the recurrence formula for the forward Green function we can work our way back to the origin, and show that the inequality holds at each intermediate node between an $x_i$ and $x_0$.  
\begin{figure}[H]
\begin{center}
\includegraphics[scale=0.7]{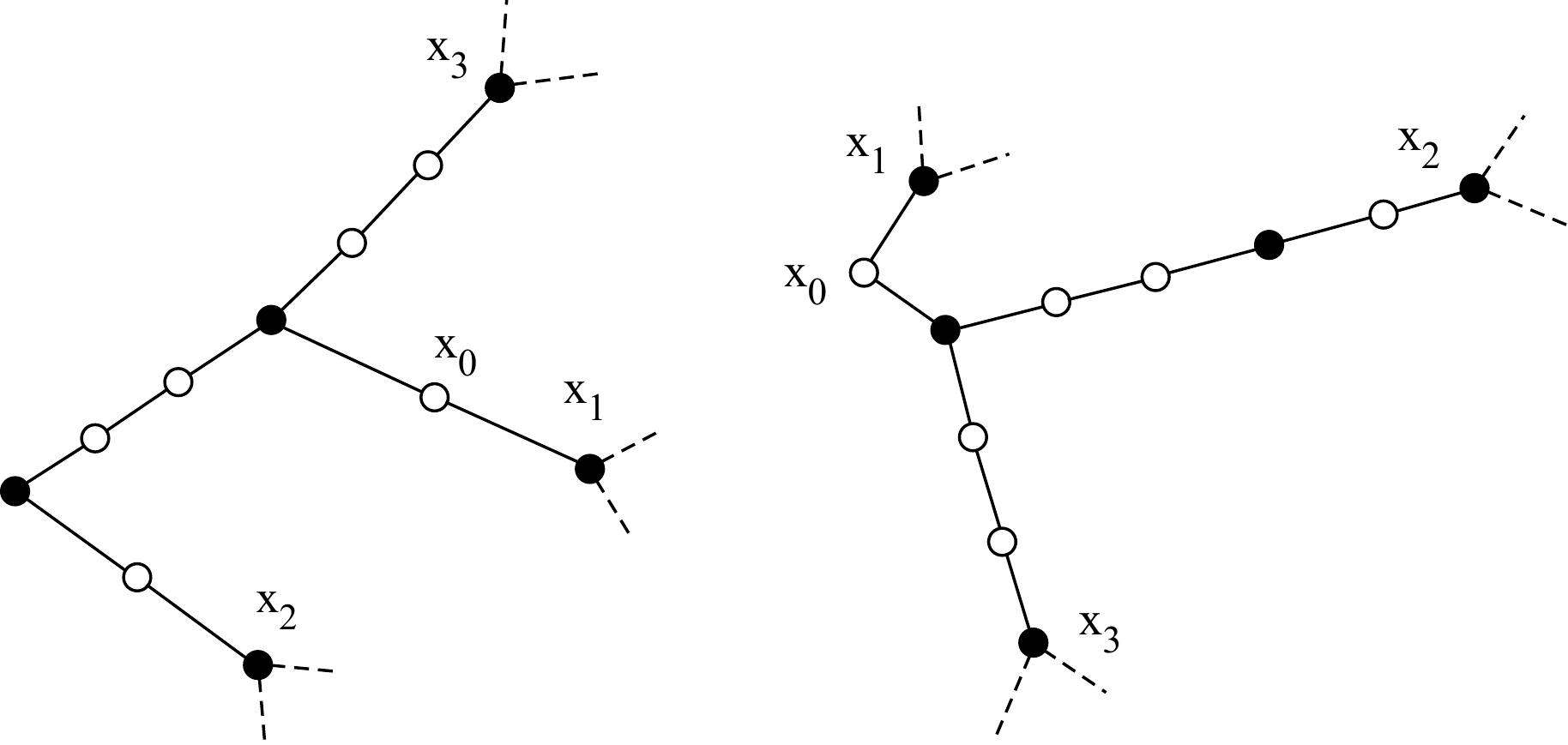}
\caption{Rearrangement of a tree.} \label{Fig.3}
\end{center}
\end{figure}

Let $y$ be such a node, forward of $x_0$ and before $x_i$. Let $\rho_j$ be the probability distribution of $G^{y_j}(\lambda)$, where $y_j$ is a neighbor of $y$ in its forward direction. We assume inductively that 
$$
\mathbb{E} \left( \mathrm{w}^{1+p}\left( \left\langle \delta_{y_j},(H^{y_j} -\lambda)^{-1}\delta_{y_j} \right\rangle\right) \right)= \int_{\mathbb{H}} \mathrm{w}^{1+p}(z_j)\, d \rho_j(z_j) < \infty\,.
$$

The functions that define the recurrence formula for the forward Green functions are fractional linear transformations and depend on the connectivity number of the node where the forward Green function is computed. 
\smallskip

\noindent {\it 1}. Assume  $y \in \mathbb{T}_a \cup \{o\}$ with $\rho'$ the probability distribution of $G^{y}(\lambda)$ and $\cal K$ a compact set in $\mathbb{H}$ such that  $z_\lambda$ is in the interior of ${\cal K}$: 
\begin{align*}
&\mathbb{E}\left(\mathrm{w}^{1+p}\left(G^y(\lambda)\right) \right) = \int_{\mathbb{H}} \mathrm{w}^{1+p}(z)\, d \rho'(z) 
= \int_{\mathbb{H}\times\mathbb{R}} \mathrm{w}^{1+p}\left(\frac{-1}{z_1+\lambda-q +1}\right)\,  d\rho_1(z_1)\, d\nu(q) \\
&= \frac{1}{2} \int_{{\cal K}^c}\left(\int_{\mathbb{R}}
\frac{\mathrm{w}^{1+p}\left(-1/(z_1+\lambda-q +1)\right)}{\mathrm{w}^{1+p}(z_1)}\,d\nu(q)\right) \times  \mathrm{w}^{1+p}(z_1) \,d\rho_1(z_1)  + C \,.
\end{align*}
The quantity $\dis \mu = \int_{\mathbb{R}} \frac{\mathrm{w}^{1+p}\left(-1/(z_1+\lambda-q +1)\right)}{\mathrm{w}^{1+p}(z_1)}\,d\nu(q)$ does not need to be less than $1$, but only bounded outside the compact set $\cal K$. Using the inequalities from the proof of Lemma \ref{L3.2}, $$\dis \mu \leq \int_{\mathbb{R}}\left(\frac{\mathrm{Im}(z_1)}{\mathrm{Im}(z_1) +\mathrm{Im}(z_\lambda)}\frac{\left( 1+|z_\lambda|^2\right)\left(1+C\left(1+|z_1|^2\right)\left(1+|k q|^2\right)\right)}{|z_1-z_\lambda|^2}\right)^{1+p}\,d\nu(q)$$ which is bounded on ${\cal K}^c$. We can therefore conclude,
\begin{align*}
&\mathbb{E}\left(\mathrm{w}^{1+p}\left(G^y(\lambda)\right) \right) \leq C' \int_{\mathbb{H}} \mathrm{w}^{1+p}(z_1)\, d \rho_1(z_1) + C = C'\, \mathbb{E}\left(\mathrm{w}^{1+p}\left(G^{y_j}(\lambda)\right) \right) + C \,.
\end{align*}
Hence  $\sup\limits_{\lambda \in R(E,\epsilon)} \mathbb{E}\left(\mathrm{w}^{1+p}\left(G^y(\lambda)\right) \right) < \infty$.
\smallskip

\noindent {\it 2}. Assume  $y \in \mathbb{T}_p$ with $\rho''$ the probability distribution of $G^{y}(\lambda)$ and  $\cal K$ a compact set in $\mathbb{H}^2$ such that  $(z_\lambda, z_\lambda)$ is in the interior of  ${\cal K}$:

\begin{align*}
&\mathbb{E}\left(\mathrm{w}^{1+p}\left(G^y(\lambda)\right) \right) = \int_{\mathbb{H}} \mathrm{w}^{1+p}(z)\, d \rho''(z) \\
&= \int_{\mathbb{H}^2\times\mathbb{R}} \mathrm{w}^{1+p}\left(\frac{-1}{z_1+z_2+\lambda-k q}\right)\,  d\rho_1(z_1)\,d\rho_2(z_2)\, d\nu(q) \\
&= \frac{1}{2} \int_{{\cal K}^c}\left(\int_{\mathbb{R}}
\frac{2\, \mathrm{w}^{1+p}\left(-1/(z_1+z_2+\lambda-q)\right)}{\mathrm{w}^{1+p}(z_1)+\mathrm{w}^{1+p}(z_2)}\,d\nu(q)\,\right) \times \big( \mathrm{w}^{1+p}(z_1)  + \\&\,\,\,\,\,\,\,\,\,\,\,\,\,\,\,\,\,\,\,\,\,\,\,\,\,\,\,\,\,\,\,\,\,\,\,\,\,\,\,\,\,\,\,\,\,\,\,\,\,\,\,\,\,\,\,\,\,\,\,\,\,\,\,\,\,\,\,\,\,\,\,\,\,\,\,\,\,\,\,\,\,\,\,\,\,\,\,\,\,\,\,\,\,\,\,\,\,\,\,\,\,\,\,\,\,\,\,\,\,\,\,\,\,\,\,\,\,+\mathrm{w}^{1+p}(z_2) \big)\,d\rho_1(z_1) \,  d\rho_2(z_2) + C \\
&\leq C'' \left(\int_{\mathbb{H}} \mathrm{w}^{1+p}(z_1)\, d \rho_1(z_1) + \int_{\mathbb{H}} \mathrm{w}^{1+p}(z_2)\, d \rho_2(z_2)\right) + C\\
&\leq C'' \left(\mathbb{E}\left(\mathrm{w}^{1+p}\left(G^{y_1}(\lambda)\right) \right)+ \mathbb{E}\left(\mathrm{w}^{1+p}\left(G^{y_2}(\lambda)\right) \right)\right)+C\,.
\end{align*}
Hence $\sup\limits_{\lambda \in R(E,\epsilon)} \mathbb{E}\left(\mathrm{w}^{1+p}\left(G^y(\lambda)\right) \right) < \infty$. The $q$ integral, outside the compact set $\cal{K}$, is bounded by arguments similar to the ones in the proof of Lemma 3. 

When we reach the origin $x_0$, we know the inequality holds at all other nodes. The recurrence relation for the origin is slightly different than everywhere else, due to our definition of the Laplacian. The argument that proves this final step is nevertheless almost identical to the one above. \qed

\section{On a recursion relation} \la{appB}

At the beginning of Section 3 we introduced quantities $A_i$, $B_i$ and $C_i$. For $q \equiv 0$, they all are recursions of the following form
 $$\ \begin{cases}
        R_{0}(z)= 1 \\
        R_1(z)=1+\lambda+z\\
        R_{n+1}(z)=(1+\lambda) R_n(z) - R_{n-1}(z)
    \end{cases}
$$
or, in a matrix form
$$\begin{bmatrix}
R_{n+1}(z)\\
R_{n}(z)
\end{bmatrix} = 
\begin{bmatrix}
1+\lambda & -1 \\
1 & 0
\end{bmatrix}
\begin{bmatrix}
R_{n}(z)\\
R_{n-1}(z)
\end{bmatrix} = \begin{bmatrix}
1+\lambda & -1 \\
1 & 0
\end{bmatrix}^{n}
\begin{bmatrix}
R_{1}(z)\\
1
\end{bmatrix}\,.
$$
 We can observe that $R_n(z)$ has the following general form, depending on $\lambda$, $R_n(z) =$ (Pol. of degree $(n-1)$ in $\lambda$) $\cdot z +$ (Pol. of degree $n$ in $\lambda$).
For $\lambda \neq -3$, $1$ we have the following diagonal form
$$\begin{bmatrix}
R_{n+1}(z)\\
R_{n}(z)
\end{bmatrix} = \frac{1}{\mathrm{det}}
\begin{bmatrix}
1 & 1 \\
\mu_2 & \mu_1
\end{bmatrix}
\begin{bmatrix}
\mu_1^n & 0 \\
0 & \mu_2^n
\end{bmatrix}\begin{bmatrix}
\mu_1 & -1 \\
-\mu_2 & 1
\end{bmatrix}
\begin{bmatrix}
R_{1}(z)\\
1
\end{bmatrix}$$
where $\dis \mu_{1,2} = \frac{1+\lambda}{2} \pm \sqrt{\left(\frac{1+\lambda}{2}\right)^2-1}$ and $\dis \mathrm{det} = 2\sqrt{\left(\frac{1+\lambda}{2}\right)^2-1}$. 

\noindent The general formula for $R_n$ is 
$\dis R_n(z) = \frac{1}{\mathrm{det}} \left( \left(\mu_1^n-\mu_2^n \right)\left(1+\lambda+z \right)- \left(\mu_1^{n-1}-\mu_2^{n-1} \right) \right)$. Also, for $n >m$ we have 
\begin{align*}
&R_n(z) = \frac{\mu_1^n - \mu_2^n}{\mu_1^m - \mu_2^m} R_m(z) + \frac{1}{\mathrm{det}} \left( \left(\mu_1^{m-1}-\mu_2^{m-1}\right) \frac{\mu_1^n - \mu_2^n}{\mu_1^m - \mu_2^m} - \left(\mu_1^{n-1}-\mu_2^{n-1}\right)\right),\\
&\hspace{0.9cm}=\frac{\mu_1^n - \mu_2^n}{\mu_1^m - \mu_2^m} R_m(z) + \frac{1}{\mathrm{det}} \frac{(-1)^{m}\left(\mu_1 - \mu_2\right)\left(\mu_1^{n-m} - \mu_2^{n-m}\right)}{\mu_1^m - \mu_2^m}.
\end{align*}

\begin{lem}\la{L3.4} The set of $\lambda$ values for which either of the following identities is true is finite:
\begin{itemize}  
\item[(i)] $R_n(z) = R_m(z) =0$;
\item[(ii)] $\dis \frac{R_n(z_\lambda)}{R_m(z_\lambda)} \in \mathbb{R}$, where $z_\lambda \in \mathbb{H}$ is the fixed point introduced in the {\rm Outline of the Proof};
\item[(iii)] $\dis \frac{R_n(z_\lambda)}{R_m(z_\lambda)} = - r\, \mathrm{i}$, where $r \in \mathbb{R}$.
\end{itemize}
\end{lem}
\proof $(i\,)$ Let us assume $R_m(z)=0$. Since we know 
\begin{align}\la{3.a1}
R_n(z)=\frac{\mu_1^n - \mu_2^n}{\mu_1^m - \mu_2^m} R_m(z) + \frac{1}{\mathrm{det}} \frac{(-1)^{m}\left(\mu_1 - \mu_2\right)\left(\mu_1^{n-m} - \mu_2^{n-m}\right)}{\mu_1^m - \mu_2^m},
\end{align}
$R_n(z) =0$ iff $\dis \frac{1}{\mathrm{det}} \frac{(-1)^{m}\left(\mu_1 - \mu_2\right)\left(\mu_1^{n-m} - \mu_2^{n-m}\right)}{\mu_1^m - \mu_2^m}=0$. This identity is equivalent to
$$
\mu_1^{n-m-1}+\mu_1^{n-m-2}\mu_2 + \ldots +\mu_1\mu_2^{n-m-2} + \mu_2^{n-m-1}=\mu_1^{m-1}+\mu_1^{m-2}\mu_2 + \ldots +\mu_1\mu_2^{m-2} + \mu_2^{m-1},
$$
which is a polynomial of degree $\max\{n-m-1, m-1\}$ in $\lambda$.

$(ii\,)$ Using (\ref{3.a1}) we can write
\begin{align}\la{3.a2}
\frac{R_n(z_\lambda)}{R_m(z_\lambda)}=\frac{\mu_1^n - \mu_2^n}{\mu_1^m - \mu_2^m} + \frac{1}{\mathrm{det}} \frac{(-1)^{m}\left(\mu_1 - \mu_2\right)\left(\mu_1^{n-m} - \mu_2^{n-m}\right)}{\left(\mu_1^m - \mu_2^m\right)R_m(z_\lambda)}.
\end{align}
The first term on the right hand side is a real number and since $R_m(z_\lambda) \notin \mathbb{R}$, the only way to obtain the desired conclusion is iff $\dis \mu_1^{n-m} - \mu_2^{n-m}=0$ which is equivalent to finding the roots of a polynomial of degree $n-m-1$ in $\lambda$.

$(iii\,)$ Relation (\ref{3.a2}) becomes
\begin{align*}
\frac{R_n(z_\lambda)}{R_m(z_\lambda)}=\frac{\mu_1^n - \mu_2^n}{\mu_1^m - \mu_2^m} + \frac{1}{\mathrm{det}} \frac{(-1)^{m}\left(\mu_1 - \mu_2\right)\left(\mu_1^{n-m} - \mu_2^{n-m}\right)\overline{R_m(z_\lambda)}}{\left(\mu_1^m - \mu_2^m\right) |R_m(z_\lambda)|^2}.
\end{align*}
For condition $(iii\,)$ to be true we need
\begin{align*}
&\frac{\mu_1^n - \mu_2^n}{\mu_1^m - \mu_2^m} + \frac{1}{\mathrm{det}} \frac{(-1)^{m}\left(\mu_1 - \mu_2\right)\left(\mu_1^{n-m} - \mu_2^{n-m}\right)\mathrm{Re}(R_m(z_\lambda))}{\left(\mu_1^m - \mu_2^m\right) |R_m(z_\lambda)|^2} = 0,
\end{align*}
which is equivalent to
\begin{align*}
&(\mu_1^n - \mu_2^n) |R_m(z_\lambda)|^2 + (-1)^{m}\left(\mu_1^{n-m} - \mu_2^{n-m}\right)\mathrm{Re}(R_m(z_\lambda)) = 0,
\end{align*}
The condition resumes to finding the zeros of a polynomial in $\lambda$. \qed


\end{document}